\documentclass[journal=jctcce,manuscript=article]{achemso}
\setkeys{acs}{maxauthors=0,articletitle=true}
\usepackage{achemso}
\usepackage{placeins}
\usepackage{graphics}
\usepackage{amssymb,amsfonts}
\usepackage{graphicx}
\usepackage[table,dvipsnames]{xcolor}
\usepackage{multirow}
\usepackage{caption}
\usepackage{subcaption}
\usepackage{booktabs}
\usepackage{colortbl}
\usepackage{amsmath}
\usepackage{amsopn}
\usepackage{bm}
\usepackage{braket}
\usepackage{siunitx}
\usepackage{color}
\usepackage{array}
\usepackage{lscape}
\usepackage{mciteplus}
\usepackage[version=3]{mhchem}
\usepackage{ulem}
\usepackage{listings}
\usepackage{enumerate}
\usepackage{lmodern}
\usepackage{mhchem}
\usepackage[implicit=false]{hyperref}

\author{Jonas Greiner}
\affiliation[mainz]{Department Chemie, Johannes Gutenberg-Universität Mainz\\Duesbergweg 10--14, 55128 Mainz, Germany}
\author{J{\"u}rgen Gauss}
\affiliation[mainz]{Department Chemie, Johannes Gutenberg-Universität Mainz\\Duesbergweg 10--14, 55128 Mainz, Germany}
\author{Janus J. Eriksen}
\email{janus@dtu.dk}
\affiliation[dtu]{DTU Chemistry, Technical University of Denmark\\Kemitorvet Bldg. 206, 2800 Kgs. Lyngby, Denmark}

\title[TITLE]{Exploiting Non-Abelian Point-Group Symmetry to Estimate the Exact Ground-State Correlation Energy of Benzene in a Polarized Split-Valence Triple-Zeta Basis Set}
\begin{document}

\begin{abstract}

Local electronic-structure methods in quantum chemistry operate on the ability to compress electron correlations more efficiently in a basis of spatially localized molecular orbitals than in a parent set of canonical orbitals. However, many typical choices of localized orbitals tend to be related by selected, near-exact symmetry operations whenever a molecule belongs to a point group, a feature which remains largely unexploited in most local correlation methods. The present Letter demonstrates how to leverage a recent unitary protocol for enforcing symmetry properties among localized orbitals to yield a high-accuracy estimate of the exact ground-state correlation energy of benzene ($D_{6h}$) in correlation-consistent polarized basis sets of both double- and triple-$\zeta$ quality. Through an initial application to many-body expanded full configuration interaction (MBE-FCI) theory, we show how molecular point-group symmetry can lead to computational savings that are inversely proportional to the order of the point group in a manner generally applicable to the acceleration of modern local correlation methods.

\end{abstract}

\vspace{1.5cm}
\begin{figure}[ht]
\begin{center}
\includegraphics[width=\textwidth]{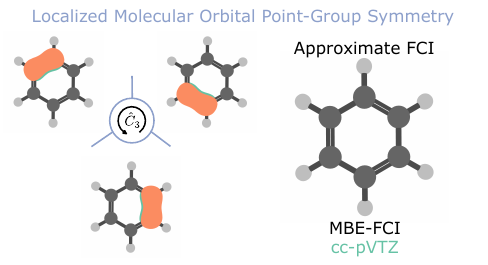}
\caption*{TOC graphic.}
\label{toc_fig}
\end{center}
\end{figure}

\newpage

Symmetry is a ubiquitous concept. Manifesting itself all around us---for instance, in the two separate halves of the human body, in leaves, snowflakes, and other objects of nature, but also in architecture, arts, and maths---symmetry may be identified everywhere and on all scales, from within the grandest entities of the universe to the smallest molecular constituents of matter. Beyond symmetry as simple mirror images, point groups generalize symmetry elements and operations, such as, reflections, inversions, and proper as well as improper rotations, and point-group arguments may even be used to predict and rationalize molecular properties and rule if certain spectroscopic transitions are allowed or not. Across the vastness of chemical compound space, the clear majority of all species will not be symmetrical in any way, but many small and modest-sized molecules of great chemical interest will be---ranging from diatomics of infinite point groups to buckminsterfullerene, the quintessential example of the icosahedral ($I_h$) point group with the largest number of possible symmetry elements.\\

The history of the exploitation of molecular point-group symmetries in quantum-chemical simulations is long and acclaimed~\cite{Pitzer1973,Davidson1975,Dupuis1977,Takada1983,Taylor1985,Taylor1986,Carsky1987,Haeser1991,Stanton1991}. The construction of a symmetry-adapted atomic orbital (AO) basis has traditionally been invoked to improve substantially upon the computational efficacy of mean-field methods. Regardless of any such symmetry adaptation of the AO basis, the resulting set of molecular orbitals (MOs) will transform as irreducible representations (irreps) of the full molecular point group, unless a symmetry-broken solution is converged upon. Consequently, correlation methods that operate in a basis of these MOs can be improved by exploiting the fact that various classes of integrals over totally symmetric operators will vanish whenever the direct products of the irreps themselves fail to contain a totally symmetric component. However, in the case of spatially localized molecular orbitals (LMOs)---obtained from a unitary rotation of a parent set of canonical molecular orbitals (CMOs) and favoured for both their ease of interpretation as well as their representational compactness---these are usually never optimized subject to the same sort of strict symmetry constraints due to the risk of suppressing their overall spatial locality as a direct and unwelcome consequence.

Be that as it may, we have recently proposed an algorithm capable of enforcing specific symmetry properties among LMOs of molecules belonging to both Abelian and higher-order point groups in a numerically exact fashion~\cite{Greiner2023}. Specifically, given how a set of symmetry-equivalent MOs will require only a single symmetry-unique orbital to span all other members of a particular subspace through the application of a befitting set of symmetry operations, this feature holds the potential to facilitate equivalences of specific integrals and related quantities across many routine quantum-chemical methods. In the present Letter, we will demonstrate how to computationally exploit such LMO redundancies---within the context of an incremental approximation to exact electronic-structure theory in a fixed AO basis---to estimate the frozen-core full configuration interaction (FCI) energy of benzene to within unprecedented accuracy in polarized split-valence basis sets of both double- and triple-$\zeta$ quality.\\

In many-body expanded full configuration interaction (MBE-FCI) theory, exact properties within a given Hilbert space are approximated without recourse to an explicit sampling of the wave function by instead performing a multitude of complete active space configuration interaction (CASCI) calculations on restricted tuples of MOs.\cite{Eriksen2017,Eriksen2018,Eriksen2019,Eriksen2019a,Eriksen2020,Eriksen2021} As was recently demonstrated in Ref.~\citenum{Greiner2024a}, MBE-FCI calculations can now be performed in an entirely unbiased manner, unlike in previous versions of the method where FCI correlation properties were decomposed and solved for by enforcing a strict partitioning of a complete set of MOs into two complementary reference and expansion spaces. Instead, an adaptive algorithm automatically identifies an optimal reference space based on quantum fidelities between individual CASCI wave functions and the Hartree-Fock (HF) determinant, upon which the residual correlation in the remaining expansion space is recovered by means of an MBE in a set of individual spatially localized MOs or suitable clusters of these. In addition, the current state of MBE-FCI operates by truncating the latter space throughout the MBE subject to an {\textit{a priori}} error tolerance, enabling more rapid and robust convergence of expanded properties onto the corresponding FCI solutions as well as resulting error bars associated with these~\cite{Greiner2024a}.

The present Letter is concerned with the exploitation of symmetry-equivalent sets of LMOs and the fact that tuples of these will give rise to identical correlation energies. For this purpose, we will make use of the \textit{petite} list method~\cite{Dupuis1977}---a technique originally intended to accelerate the evaluation of integrals---to recover symmetry equivalences among MBE increments. By looping over all orbital combinations (the \textit{grande} list) and applying the symmetry operations of the molecular point group to determine whether a given orbital tuple is lexicographically greater than any other symmetry-equivalent orbital tuple, it is possible to single out which tuple of MOs for a set of symmetry-equivalent active spaces will need to be explicitly considered. These symmetry-unique orbital combinations will then form the \textit{petite} list. It is important to emphasize how our proposed use of symmetrized LMOs ({\textit{vide infra}}) is generally applicable to a manifold of local electron-correlation schemes whenever these can formally express the correlation energy as a sum over contributions from individual MOs.\\

Our generalization of \textit{petite} lists to arbitrary orbital tensors in an MBE requires that these objects must be subject to numerically exact symmetries. Due to the fact that cost functions of different orbital localization schemes possess numerous, densely clustered local minima, subject only to constraints on spatial locality, LMOs must be explicitly symmetrized as in Ref. \citenum{Greiner2023}. This procedure is of particular importance in the context of applications to orbital-based MBEs, given the recursive nature of such methods and the fact that any errors related to near-equivalences will propagate and become amplified throughout an expansion~\cite{Friedrich2008}.\\

\begin{figure}[ht!]
    \includegraphics[width=\textwidth]{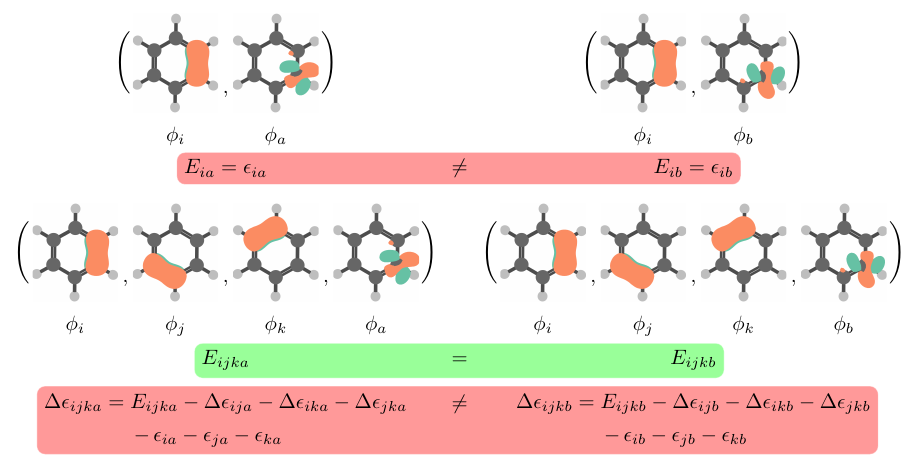}
    \caption{Equivalences and differences in correlation ($E$) and increment ($\epsilon,\Delta\epsilon$) energies.}
    \label{symm_eqv_incs}
\end{figure}

Our implementation of symmetry within MBE-FCI theory differentiates between symmetry equivalences of {\textit{increments}} and {\textit{active spaces}}. A set of symmetry-equivalent active spaces all produce the same CASCI property but will only produce symmetry-equivalent increments in case this equivalence holds true for all subtuple active spaces. Thus, while a parent tuple may be symmetry-equivalent, its individual MOs may risk being neither equivalent nor invariant with respect to a given symmetry operation. The selected Pipek-Mezey~\cite{Pipek1989} (PM) LMOs of benzene in Fig.~\ref{symm_eqv_incs} provides an illustrative example of exactly this situation.

The upper panel of Fig.~\ref{symm_eqv_incs} depicts two orbital tuples and how their CASCI correlation energies ($\{E_{ia},E_{ib}\}$) and increments ($\{\epsilon_{ia},\epsilon_{ib}\}$) are different. Both tuples involve the same occupied $\pi$-orbital, $\phi_i$, and two different virtual orbitals, $\phi_a$ and $\phi_b$, from a set of 12 symmetry-equivalent orbitals. While these two virtual orbitals can be transformed into each other by applying a vertical mirror plane, $\phi_i$, on the other hand, is neither symmetry-invariant nor equivalent with any other $\pi$-orbital in the symmetry-equivalent set of cardinality 3 when applying said symmetry operation. On the contrary, the bottom panel of Fig.~\ref{symm_eqv_incs} depicts active spaces formed from the complete set of occupied $\pi$-orbitals and the same two virtual orbitals, $\{\phi_a,\phi_b\}$. While the CASCI correlation energies are equal due to the symmetry-invariance of the $\pi$-orbital space with respect to all symmetry operations, the corresponding recursive increments will not be equal. In our implementation, whenever orbital tuples fail to produce symmetry-equivalent increments, the CASCI calculation in question is performed only once before the required increments get calculated from appropriate contributions only. The supporting information (SI) provides further details on our parallel implementation~\cite{pymbe}.

\begin{figure}[ht!]
    \includegraphics[width=0.75\textwidth]{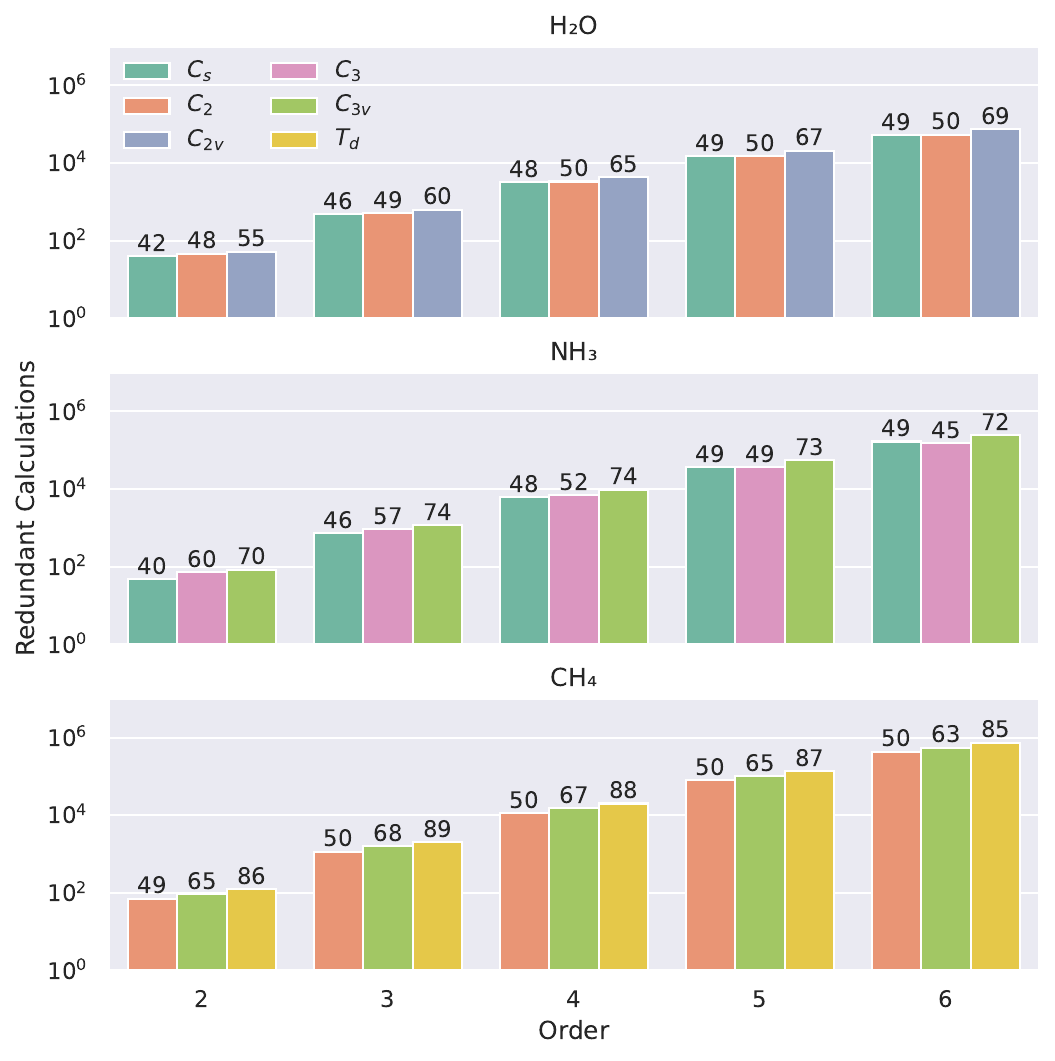}
    \caption{Total number (and relative percentage) of redundant calculations for specific point groups plotted against the MBE order for H$_2$O, NH$_3$, and CH$_4$ in a cc-pVDZ basis. These all-electron calculations use symmetrized FB LMOs and are based on empty reference spaces.}
    \label{mol_symm_empty}
\end{figure}
Our proposed symmetry exploitation is next evaluated by performing MBE-FCI calculations on water ($C_{2v}$), ammonia ($C_{3v}$), and methane ($T_d$) in the standard cc-pVDZ basis set~\cite{Dunning1989}. The results in Fig. \ref{mol_symm_empty}---based on Foster-Boys~\cite{Foster1960} (FB) orbitals---show how both the total and redundant number of increment calculations increase exponentially, ensuring that the relative number of increment calculations to skip stays reasonably constant as the MBE order rises. The actual percentage is highly dependent on both the chosen point group of the molecular system and the nature of the LMOs themselves, as was previously touched upon in Ref. \citenum{Greiner2023}. The upper bound for symmetry exploitation is inversely proportional to the order of the molecular point group, $h$, an asymptotic limit which will only be reached when all MOs form symmetry-equivalent sets whose cardinality is equal to $h$. Given the limited size of all three molecular examples in Fig.~\ref{mol_symm_empty}, these upper bounds are never formally met since a few orbitals will always stay fully localized around the central (O, N, or C) atom.

\begin{figure}[ht!]
    \includegraphics[width=0.75\textwidth]{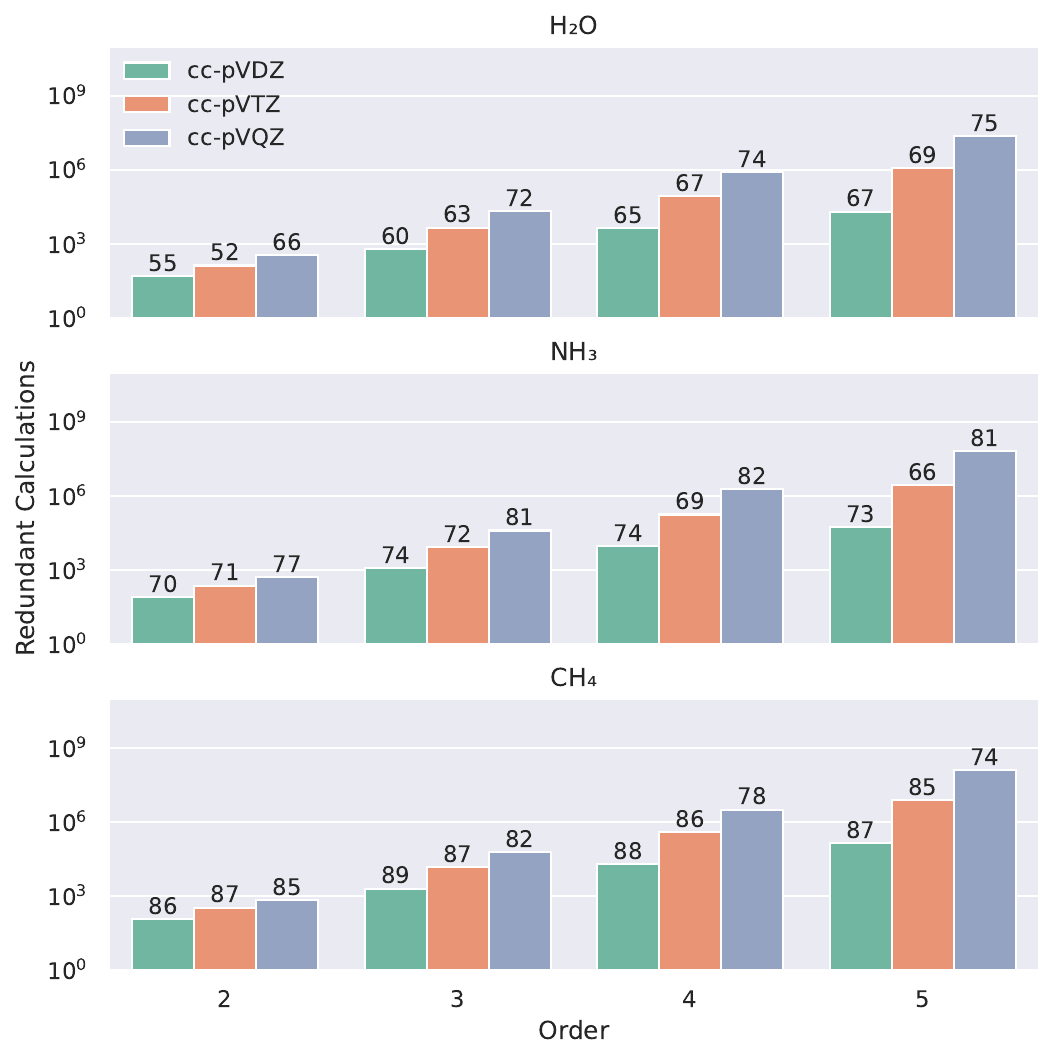}
    \caption{Total number (and relative percentage) of redundant calculations in the cc-pV$X$Z basis sets for H$_2$O, NH$_3$, and CH$_4$ in their respective full molecular point groups. These all-electron calculations use symmetrized FB LMOs and are based on empty reference spaces.}
    \label{mol_symm_basis_empty}
\end{figure}
By repeating the calculations of Fig. \ref{mol_symm_empty} in the larger cc-pVTZ and cc-pVQZ basis sets, the results in Fig.~\ref{mol_symm_basis_empty} demonstrate how the relative number of redundant calculations stays largely invariant under a change of the one-electron basis. However, in the case of extended basis sets, the localization cost function for the virtual orbital space may suffer from an abundantly large number of local minima~\cite{Subotnik2004,Subotnik2005}, and the symmetry properties of the converged minimum can adversely affect the ability to exploit point-group symmetry, as can be seen for NH$_3$ in the cc-pVTZ basis and CH$_4$ in the cc-pVQZ basis. On the other hand, the opposite may also hold true, as is clearly evident in the results for H$_2$O and NH$_3$ in the cc-pVQZ basis.

\begin{figure}[ht!]
    \includegraphics[width=0.85\textwidth]{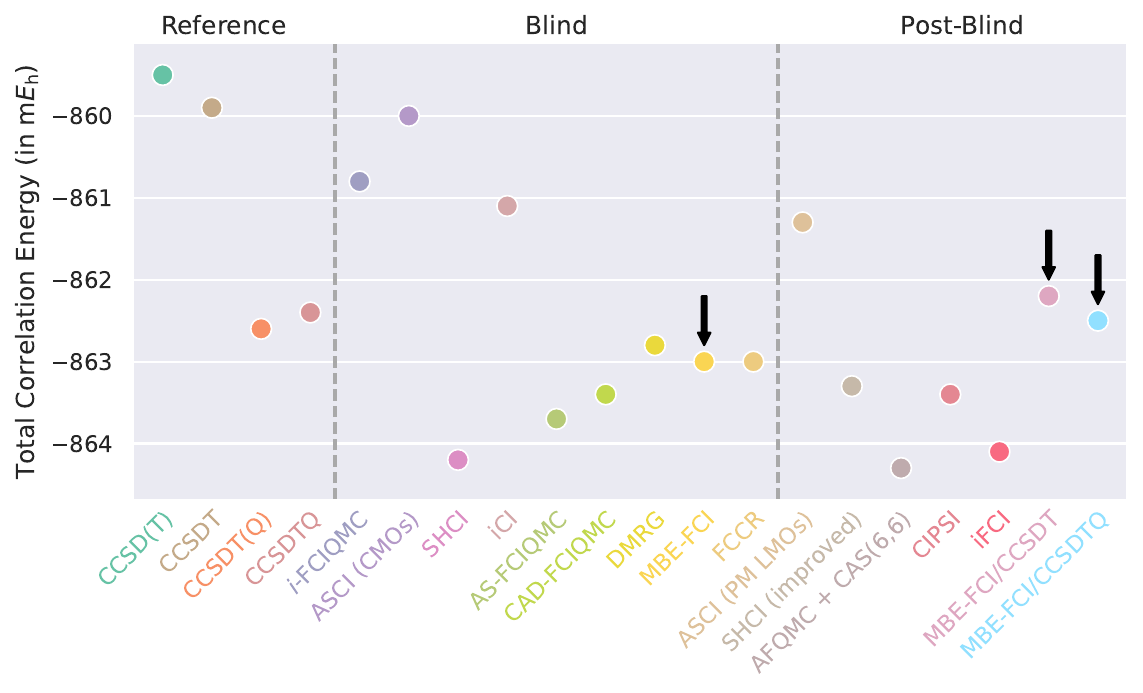}
    \caption{Results for the frozen-core ground-state correlation energy of benzene in the cc-pVDZ basis set. The $i$-FCIQMC result is taken from Ref.~\citenum{Blunt2019}, and the remaining blind results and post-blind results for ASCI and SHCI are taken from Ref.~\citenum{Eriksen2020a}. The AFQMC, CIPSI, and iFCI results are taken from Refs.~\citenum{Lee2020}, \citenum{Loos2020}, and \citenum{Rask2021}, respectively. A comprehensive list of the acronyms behind all the entries here is provided in Table S1 at the start of the SI.}
    \label{c6h6_dz_benchmark}
\end{figure}
A popular symbol of the chemical sciences, the benzene molecule was recently the subject of a large, international blind challenge devoted to determining its ground-state (frozen-core) electronic energy in a modest correlation-consistent polarized double-$\zeta$ basis (cc-pVDZ) using a wide range of contemporary, high-accuracy approximations to FCI~\cite{Eriksen2020a}. Following its publication, additional (post-blind) results have been reported. Among the entries in Ref. \citenum{Eriksen2020a}, the MBE-FCI method was able to converge the correlation energy of the benzene system to $-863.0 \ \mathrm{m}E_\mathrm{h}$ by employing a reference space of 6 localized PM $\pi$-orbitals and a corresponding single-orbital expansion space, an estimated value situated close to the most dense part of the energy spectrum predicted by the suite of approximate FCI methods, cf. Fig. \ref{c6h6_dz_benchmark}.\\

Alas, MBE-FCI was also by far the most expensive method benchmarked in Ref. \citenum{Eriksen2020a}, demanding a staggering 1.7 million core hours~\bibnote{Computer architecture: Intel Xeon E5-2697v4 (Broadwell) hardware (36 cores @ 2.3 GHz, 3.56 GB/core).}. One key reason for the large discrepancy in computational cost compared to other methods is the fact that MBE-FCI inherently involves a substantial amount of redundant increment calculations when applied to molecules of high symmetry. As an integral part of the present Letter, we will demonstrate how to lower the computational cost associated with running MBE-FCI calculations, not just to bring the method to a level on par with other modern FCI approximations, but importantly also to be able to enlarge the one-electron basis to one of polarized split-valence triple-$\zeta$ quality (cc-pVTZ). Beyond the application of improved screening protocols, tailored reference spaces, and clustered expansion spaces~\cite{Greiner2024a}, the proper exploitation of point-group symmetry in an optimized LMO basis is of paramount importance for achieving this objective. In addition, to further improve upon the accuracy of MBE-FCI, we will opt to target the gap in energy between FCI and a so-called base model (cf. Ref. \citenum{Eriksen2017}) rather than the full correlation energy~\cite{Eriksen2020a}.\\

The SI contains a detailed discussion on how optimal settings behind our MBE-FCI calculations were chosen. In brief summary, a combination of a reference space consisting of the two sets of degenerate $\pi$-orbitals in the CMO basis and an expansion space of PM LMOs was found to allow for correlation to be optimally accounted for in the former space, while still permitting tight convergence of an MBE in the latter space upon moving to high expansion orders. Likewise, MO pairs rather than single MOs as the expansion objects were found to be beneficial in terms of reductions to the number of increment calculations that need to be evaluated. Finally, coupled cluster models with a full account of triple~\cite{ccsdt_paper_1_jcp_1987,ccsdt_paper_1_jcp_1988_erratum,ccsdt_paper_2_cpl_1988} (CCSDT) and quadruple~\cite{ccsdtq_paper_1_jcp_1991,ccsdtq_paper_2_jcp_1992} (CCSDTQ) excitations were employed as so-called base models, given how both are expected to constitute favourable and orbital-invariant approximations to FCI, unlike, for instance, the perturbative CCSD(T)~\cite{original_ccsdpt_paper} and CCSDT(Q)~\cite{bomble_ccsdt_pq_jcp_2005} models, cf. Fig. \ref{c6h6_dz_benchmark}.\\

The results of MBE-FCI calculations on benzene in the cc-pVDZ basis---in full $D_{6h}$ point-group symmetry and employing both choices of base models---are compared to other state-of-the-art methods in Fig. \ref{c6h6_dz_benchmark}. While most other post-blind results for the correlation energy fall below the average result of Ref. \citenum{Eriksen2020a}, both energies predicted in the present Letter are instead shifted upwards, in perfect agreement with the CCSDTQ reference result, the residual error of which is expected to be marginal with respect to FCI. While the overall interpretation of both accuracy as well as reliability of all the individual entries in Fig. \ref{c6h6_dz_benchmark} is left to the reader, our updated MBE-FCI results undoubtedly constitute a significant improvement upon the original result in Ref. \citenum{Eriksen2020a}, not merely in terms of accuracy (on account of our use of base models), but importantly also in terms of time-to-solution, due to orbital clustering and, foremost, the efficient exploitation of point-group symmetry. The MBE-FCI/CCSDT and MBE-FCI/CCSDTQ calculations required no more than 1,700 and 2,600 core hours, respectively~\bibnote{Computer architecture used throughout the present study: Two Intel Gold 6130 per node (32 cores @ 2.1 GHz, 5.53 GB/core). The initial base model calculation on the full system has not been included in all reported timings as this preliminary step is not amenable to embarrassingly parallel computing in the same way as MBE-FCI is.}; in comparison to the MBE-FCI calculation reported in Ref. \citenum{Eriksen2020a}, the computational effort has thus been reduced by a massive three orders of magnitude. In accumulative figures, a total of $89\%$ of the required CASCI calculations were omitted due to our handling of point-group symmetries among the LMOs (2.6M redundant calculations).

\begin{figure}[ht!]
    \includegraphics[width=0.85\textwidth]{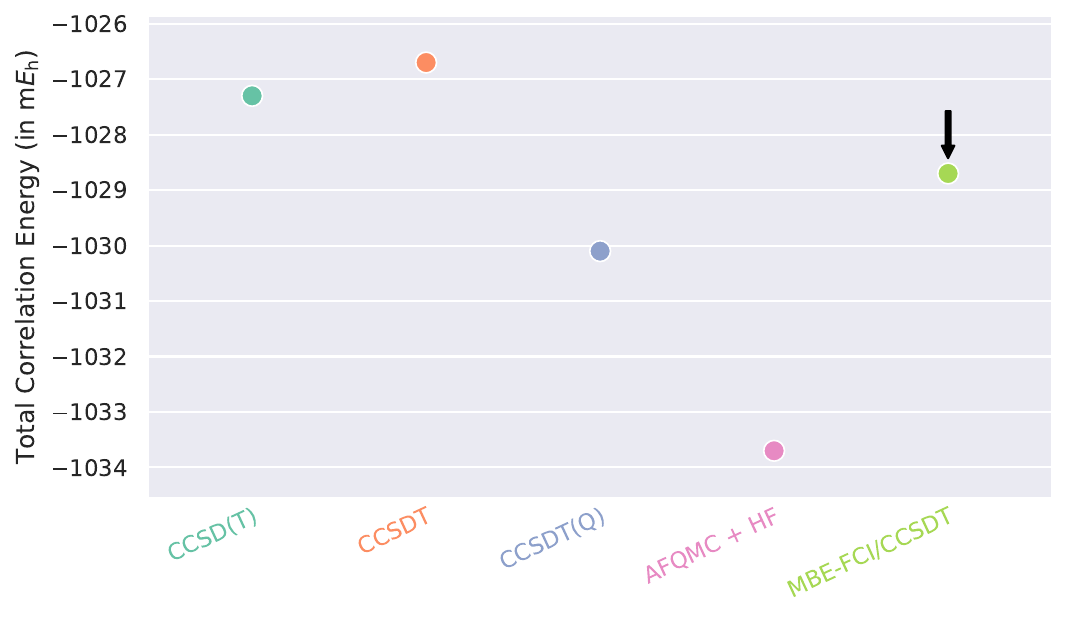}
    \caption{Results for the frozen-core ground-state correlation energy of benzene in the cc-pVTZ basis set. The HF-based AFQMC result is taken from Ref. \citenum{Lee2020}.}
    \label{c6h6_tz_benchmark}
\end{figure}
Having reassessed the performance of the revamped MBE-FCI method in the modest cc-pVDZ basis, we will now extend the description of the benzene system to the larger cc-pVTZ basis set. While a CCSDTQ base model calculation is infeasible for a system with $N = 254$ orbitals to correlate, given its formal $\mathcal{O}(N^{10})$ scaling, CCSDT can still be employed as a base model. Fig. \ref{c6h6_tz_benchmark} compares the CCSDT reference result to alternative CC methods as well as a quantum Monte Carlo result from Ref. \citenum{Lee2020}. Although not variational, it is arguably fair to assume that the CCSDT result is too high in energy for this system, also considering the comparison between CCSDT and CCSDTQ in the smaller cc-pVDZ basis, cf. Fig. \ref{c6h6_dz_benchmark}. The perturbative CCSDT(Q) model, on the other hand, likely yields an energy slightly below the exact answer, again drawing conclusions on the basis of Fig. \ref{c6h6_dz_benchmark}, but also its known tendency to overshoot correlation energies of the full CCSDTQ method~\cite{eriksen_ccsd_pert_theory_jcp_2014,eriksen_convergence_ccpt_jcp_2016,eriksen_quadruples_pert_theory_jcp_2015,eriksen_open_shell_quadruples_jcp_2016}. The MBE-FCI/CCSDT entry in Fig. \ref{c6h6_tz_benchmark} yields a correlation energy of $-1028.7 \ mE_\mathrm{h}$, observed to fall right between the CCSDT and CCSDT(Q) result. While this energy will likely decrease slightly at potential later orders of the MBE, comparisons with the more accurate MBE-FCI result employing the CCSDTQ base model in the cc-pVDZ basis (Fig. \ref{c6h6_dz_benchmark}) indicate that the resulting error will not exceed a conservative estimate of 1 kcal/mol. On account of our utilization of the full $D_{6h}$ point group, a massive 238M CASCI calculations (90\%) could be skipped due to symmetry equivalences, and the calculation completed in only a total of 39,700 core hours.\\

In summary, we have introduced a novel method for exploiting (non-)Abelian point-group symmetries in electronic-structure methods that operate in a basis of spatially localized molecular orbitals. On account of an exact symmetrization of said orbital basis, equivalences of tensor elements or incremental energy contributions, as in the present application to MBE-FCI theory, will lead to significant computational savings, which we have verified through numerical tests using various basis sets for a number of systems belonging to different molecular point groups. In combination with an efficient clustering of orbitals, the developments reported here have been demonstrated to considerably improve the computational efficacy of MBE-FCI theory. In further conjunction with correlated base models, we have applied MBE-FCI to the calculation of ground-state correlation energies of benzene in polarized split-valence basis sets of both double- and triple-$\zeta$ quality. Leveraging the full $D_{6h}$ point group of the benzene system, our earlier cc-pVDZ result from Ref. \citenum{Eriksen2020a} has been substantially refined at a significantly lower cost, and results in the larger cc-pVTZ basis have been made feasible in convincing agreement with high-level coupled cluster results. On account of the efficient exploitation of point-group symmetries reported herein, we believe MBE-FCI can help pave the way towards near-exact calculations on molecules and intermolecular complexes of unprecedented sizes. In addition, the integrated use of symmetry within incremental formulations of complete active space self-consistent field (CASSCF) methods holds promise of extending the scope of systems amenable to such electronic-structure treatments~\cite{Greiner2024}.

\section*{Acknowledgments}

This work was supported by two research grants awarded to JJE, no. 37411 from VILLUM FONDEN (a part of THE VELUX FOUNDATIONS) and no. 10.46540/2064-00007B from the Independent Research Fund Denmark. The authors gratefully acknowledge the computing time granted on the Mogon II supercomputer by the Johannes Gutenberg-Universität Mainz ({\url{hpc.uni-mainz.de}}).

\section*{Supporting Information}

The supporting information (SI) contains a list of all acronyms of Figs. \ref{c6h6_dz_benchmark} and \ref{c6h6_tz_benchmark} (Table S1), further details on our implementation of symmetry within MBE-FCI and the {\texttt{PyMBE}} code, as well as details on the calibration of the MBE-FCI calculations behind Figs. \ref{c6h6_dz_benchmark} and \ref{c6h6_tz_benchmark}.

\section*{Data Availability}

Data in support of the findings of this study are available within the article and its SI.

\newpage

\providecommand{\latin}[1]{#1}
\makeatletter
\providecommand{\doi}
  {\begingroup\let\do\@makeother\dospecials
  \catcode`\{=1 \catcode`\}=2 \doi@aux}
\providecommand{\doi@aux}[1]{\endgroup\texttt{#1}}
\makeatother
\providecommand*\mcitethebibliography{\thebibliography}
\csname @ifundefined\endcsname{endmcitethebibliography}
  {\let\endmcitethebibliography\endthebibliography}{}


\begin{mcitethebibliography}{44}
\providecommand*\natexlab[1]{#1}
\providecommand*\mciteSetBstSublistMode[1]{}
\providecommand*\mciteSetBstMaxWidthForm[2]{}
\providecommand*\mciteBstWouldAddEndPuncttrue
  {\def\EndOfBibitem{\unskip.}}
\providecommand*\mciteBstWouldAddEndPunctfalse
  {\let\EndOfBibitem\relax}
\providecommand*\mciteSetBstMidEndSepPunct[3]{}
\providecommand*\mciteSetBstSublistLabelBeginEnd[3]{}
\providecommand*\EndOfBibitem{}
\mciteSetBstSublistMode{f}
\mciteSetBstMaxWidthForm{subitem}{(\alph{mcitesubitemcount})}
\mciteSetBstSublistLabelBeginEnd
  {\mcitemaxwidthsubitemform\space}
  {\relax}
  {\relax}

\bibitem[Pitzer(1973)]{Pitzer1973}
Pitzer,~R.~M. Electron Repulsion Integrals and Symmetry Adapted Charge
  Distributions. \emph{J. Chem. Phys.} \textbf{1973}, \emph{59}, 3308\relax
\mciteBstWouldAddEndPuncttrue
\mciteSetBstMidEndSepPunct{\mcitedefaultmidpunct}
{\mcitedefaultendpunct}{\mcitedefaultseppunct}\relax
\EndOfBibitem
\bibitem[Davidson(1975)]{Davidson1975}
Davidson,~E.~R. Use of Double Cosets in Constructing Integrals Over Symmetry
  Orbitals. \emph{J. Chem. Phys.} \textbf{1975}, \emph{62}, 400\relax
\mciteBstWouldAddEndPuncttrue
\mciteSetBstMidEndSepPunct{\mcitedefaultmidpunct}
{\mcitedefaultendpunct}{\mcitedefaultseppunct}\relax
\EndOfBibitem
\bibitem[Dupuis and King(1977)Dupuis, and King]{Dupuis1977}
Dupuis,~M.; King,~H.~F. Molecular Symmetry and Closed-Shell {SCF} Calculations.
  I. \emph{Int. J. Quantum Chem.} \textbf{1977}, \emph{11}, 613\relax
\mciteBstWouldAddEndPuncttrue
\mciteSetBstMidEndSepPunct{\mcitedefaultmidpunct}
{\mcitedefaultendpunct}{\mcitedefaultseppunct}\relax
\EndOfBibitem
\bibitem[Takada \latin{et~al.}(1983)Takada, Dupuis, and King]{Takada1983}
Takada,~T.; Dupuis,~M.; King,~H.~F. Molecular Symmetry. {IV}. The Coupled
  Perturbed Hartree–Fock Method. \emph{J. Comput. Chem.} \textbf{1983},
  \emph{4}, 234\relax
\mciteBstWouldAddEndPuncttrue
\mciteSetBstMidEndSepPunct{\mcitedefaultmidpunct}
{\mcitedefaultendpunct}{\mcitedefaultseppunct}\relax
\EndOfBibitem
\bibitem[Taylor(1985)]{Taylor1985}
Taylor,~P.~R. Symmetrization of Operator Matrix Elements. \emph{Int. J. Quantum
  Chem.} \textbf{1985}, \emph{27}, 89\relax
\mciteBstWouldAddEndPuncttrue
\mciteSetBstMidEndSepPunct{\mcitedefaultmidpunct}
{\mcitedefaultendpunct}{\mcitedefaultseppunct}\relax
\EndOfBibitem
\bibitem[Taylor(1986)]{Taylor1986}
Taylor,~P.~R. Symmetry-Adapted Integral Derivatives. \emph{Theor. Chim. Acta}
  \textbf{1986}, \emph{69}, 447\relax
\mciteBstWouldAddEndPuncttrue
\mciteSetBstMidEndSepPunct{\mcitedefaultmidpunct}
{\mcitedefaultendpunct}{\mcitedefaultseppunct}\relax
\EndOfBibitem
\bibitem[Čársky \latin{et~al.}(1987)Čársky, Schaad, Hess, Urban, and
  Noga]{Carsky1987}
Čársky,~P.; Schaad,~L.~J.; Hess,~B.~A.; Urban,~M.; Noga,~J. Use of Molecular
  Symmetry in Coupled-Cluster Theory. \emph{J. Chem. Phys.} \textbf{1987},
  \emph{87}, 411\relax
\mciteBstWouldAddEndPuncttrue
\mciteSetBstMidEndSepPunct{\mcitedefaultmidpunct}
{\mcitedefaultendpunct}{\mcitedefaultseppunct}\relax
\EndOfBibitem
\bibitem[Häser \latin{et~al.}(1991)Häser, Almlöf, and
  Feyereisen]{Haeser1991}
Häser,~M.; Almlöf,~J.; Feyereisen,~M.~W. Exploiting Non-Abelian Point Group
  Symmetry in Direct Two-Electron Integral Transformations. \emph{Theor. Chim.
  Acta} \textbf{1991}, \emph{79}, 115\relax
\mciteBstWouldAddEndPuncttrue
\mciteSetBstMidEndSepPunct{\mcitedefaultmidpunct}
{\mcitedefaultendpunct}{\mcitedefaultseppunct}\relax
\EndOfBibitem
\bibitem[Stanton \latin{et~al.}(1991)Stanton, Gauss, Watts, and
  Bartlett]{Stanton1991}
Stanton,~J.~F.; Gauss,~J.; Watts,~J.~D.; Bartlett,~R.~J. A Direct Product
  Decomposition Approach for Symmetry Exploitation in Many-Body Methods. I.
  Energy Calculations. \emph{J. Chem. Phys.} \textbf{1991}, \emph{94},
  4334\relax
\mciteBstWouldAddEndPuncttrue
\mciteSetBstMidEndSepPunct{\mcitedefaultmidpunct}
{\mcitedefaultendpunct}{\mcitedefaultseppunct}\relax
\EndOfBibitem
\bibitem[Greiner and Eriksen(2023)Greiner, and Eriksen]{Greiner2023}
Greiner,~J.; Eriksen,~J.~J. Symmetrization of Localized Molecular Orbitals.
  \emph{J. Phys. Chem. A} \textbf{2023}, \emph{127}, 3535\relax
\mciteBstWouldAddEndPuncttrue
\mciteSetBstMidEndSepPunct{\mcitedefaultmidpunct}
{\mcitedefaultendpunct}{\mcitedefaultseppunct}\relax
\EndOfBibitem
\bibitem[Eriksen \latin{et~al.}(2017)Eriksen, Lipparini, and
  Gauss]{Eriksen2017}
Eriksen,~J.~J.; Lipparini,~F.; Gauss,~J. Virtual Orbital Many-Body Expansions:
  A Possible Route Towards the Full Configuration Interaction Limit. \emph{J.
  Phys. Chem. Lett.} \textbf{2017}, \emph{8}, 4633\relax
\mciteBstWouldAddEndPuncttrue
\mciteSetBstMidEndSepPunct{\mcitedefaultmidpunct}
{\mcitedefaultendpunct}{\mcitedefaultseppunct}\relax
\EndOfBibitem
\bibitem[Eriksen and Gauss(2018)Eriksen, and Gauss]{Eriksen2018}
Eriksen,~J.~J.; Gauss,~J. Many-Body Expanded Full Configuration Interaction. I.
  Weakly Correlated Regime. \emph{J. Chem. Theory Comput.} \textbf{2018},
  \emph{14}, 5180\relax
\mciteBstWouldAddEndPuncttrue
\mciteSetBstMidEndSepPunct{\mcitedefaultmidpunct}
{\mcitedefaultendpunct}{\mcitedefaultseppunct}\relax
\EndOfBibitem
\bibitem[Eriksen and Gauss(2019)Eriksen, and Gauss]{Eriksen2019}
Eriksen,~J.~J.; Gauss,~J. Many-Body Expanded Full Configuration Interaction.
  {II}. Strongly Correlated Regime. \emph{J. Chem. Theory Comput.}
  \textbf{2019}, \emph{15}, 4873\relax
\mciteBstWouldAddEndPuncttrue
\mciteSetBstMidEndSepPunct{\mcitedefaultmidpunct}
{\mcitedefaultendpunct}{\mcitedefaultseppunct}\relax
\EndOfBibitem
\bibitem[Eriksen and Gauss(2019)Eriksen, and Gauss]{Eriksen2019a}
Eriksen,~J.~J.; Gauss,~J. Generalized Many-Body Expanded Full Configuration
  Interaction Theory. \emph{J. Phys. Chem. Lett.} \textbf{2019}, \emph{10},
  7910\relax
\mciteBstWouldAddEndPuncttrue
\mciteSetBstMidEndSepPunct{\mcitedefaultmidpunct}
{\mcitedefaultendpunct}{\mcitedefaultseppunct}\relax
\EndOfBibitem
\bibitem[Eriksen and Gauss(2020)Eriksen, and Gauss]{Eriksen2020}
Eriksen,~J.~J.; Gauss,~J. Ground and Excited State First-Order Properties in
  Many-Body Expanded Full Configuration Interaction Theory. \emph{J. Chem.
  Phys.} \textbf{2020}, \emph{153}, 154107\relax
\mciteBstWouldAddEndPuncttrue
\mciteSetBstMidEndSepPunct{\mcitedefaultmidpunct}
{\mcitedefaultendpunct}{\mcitedefaultseppunct}\relax
\EndOfBibitem
\bibitem[Eriksen and Gauss(2021)Eriksen, and Gauss]{Eriksen2021}
Eriksen,~J.~J.; Gauss,~J. Incremental Treatments of the Full Configuration
  Interaction Problem. \emph{Wiley Interdiscip. Rev.: Comput. Mol. Sci.}
  \textbf{2021}, \emph{11}, e1525\relax
\mciteBstWouldAddEndPuncttrue
\mciteSetBstMidEndSepPunct{\mcitedefaultmidpunct}
{\mcitedefaultendpunct}{\mcitedefaultseppunct}\relax
\EndOfBibitem
\bibitem[Greiner \latin{et~al.}(2024)Greiner, Gauss, and Eriksen]{Greiner2024a}
Greiner,~J.; Gauss,~J.; Eriksen,~J.~J. {Error Control and Automatic Detection
  of Reference Active Spaces in Many-Body Expanded Full Configuration
  Interaction}. 2024; arXiv:2406.11343\relax
\mciteBstWouldAddEndPuncttrue
\mciteSetBstMidEndSepPunct{\mcitedefaultmidpunct}
{\mcitedefaultendpunct}{\mcitedefaultseppunct}\relax
\EndOfBibitem
\bibitem[Friedrich \latin{et~al.}(2008)Friedrich, Hanrath, and
  Dolg]{Friedrich2008}
Friedrich,~J.; Hanrath,~M.; Dolg,~M. Using Symmetry in the Framework of the
  Incremental Scheme: Molecular Applications. \emph{Chem. Phys.} \textbf{2008},
  \emph{346}, 266\relax
\mciteBstWouldAddEndPuncttrue
\mciteSetBstMidEndSepPunct{\mcitedefaultmidpunct}
{\mcitedefaultendpunct}{\mcitedefaultseppunct}\relax
\EndOfBibitem
\bibitem[Pipek and Mezey(1989)Pipek, and Mezey]{Pipek1989}
Pipek,~J.; Mezey,~P.~G. A Fast Intrinsic Localization Procedure Applicable for
  \textit{Ab Initio} and Semiempirical Linear Combination of Atomic Orbital
  Wave Functions. \emph{J. Chem. Phys.} \textbf{1989}, \emph{90}, 4916\relax
\mciteBstWouldAddEndPuncttrue
\mciteSetBstMidEndSepPunct{\mcitedefaultmidpunct}
{\mcitedefaultendpunct}{\mcitedefaultseppunct}\relax
\EndOfBibitem
\bibitem[Greiner and Eriksen()Greiner, and Eriksen]{pymbe}
Greiner,~J.; Eriksen,~J.~J. {{\texttt{PyMBE}}: A Many-Body Expanded Correlation
  Code}. See: {\url{https://gitlab.com/januseriksen/pymbe}}\relax
\mciteBstWouldAddEndPuncttrue
\mciteSetBstMidEndSepPunct{\mcitedefaultmidpunct}
{\mcitedefaultendpunct}{\mcitedefaultseppunct}\relax
\EndOfBibitem
\bibitem[Dunning(1989)]{Dunning1989}
Dunning,~T.~H. Gaussian Basis Sets for Use in Correlated Molecular
  Calculations. {I}. The Atoms Boron through Neon and Hydrogen. \emph{J. Chem.
  Phys.} \textbf{1989}, \emph{90}, 1007\relax
\mciteBstWouldAddEndPuncttrue
\mciteSetBstMidEndSepPunct{\mcitedefaultmidpunct}
{\mcitedefaultendpunct}{\mcitedefaultseppunct}\relax
\EndOfBibitem
\bibitem[Foster and Boys(1960)Foster, and Boys]{Foster1960}
Foster,~J.~M.; Boys,~S.~F. Canonical Configurational Interaction Procedure.
  \emph{Rev. Mod. Phys} \textbf{1960}, \emph{32}, 300\relax
\mciteBstWouldAddEndPuncttrue
\mciteSetBstMidEndSepPunct{\mcitedefaultmidpunct}
{\mcitedefaultendpunct}{\mcitedefaultseppunct}\relax
\EndOfBibitem
\bibitem[Subotnik \latin{et~al.}(2004)Subotnik, Shao, Liang, and
  Head-Gordon]{Subotnik2004}
Subotnik,~J.~E.; Shao,~Y.; Liang,~W.; Head-Gordon,~M. An Efficient Method for
  Calculating Maxima of Homogeneous Functions of Orthogonal Matrices:
  Applications to Localized Occupied Orbitals. \emph{J. Chem. Phys.}
  \textbf{2004}, \emph{121}, 9220\relax
\mciteBstWouldAddEndPuncttrue
\mciteSetBstMidEndSepPunct{\mcitedefaultmidpunct}
{\mcitedefaultendpunct}{\mcitedefaultseppunct}\relax
\EndOfBibitem
\bibitem[Subotnik \latin{et~al.}(2005)Subotnik, Dutoi, and
  Head-Gordon]{Subotnik2005}
Subotnik,~J.~E.; Dutoi,~A.~D.; Head-Gordon,~M. Fast Localized Orthonormal
  Virtual Orbitals Which Depend Smoothly on Nuclear Coordinates. \emph{J. Chem.
  Phys.} \textbf{2005}, \emph{123}, 114108\relax
\mciteBstWouldAddEndPuncttrue
\mciteSetBstMidEndSepPunct{\mcitedefaultmidpunct}
{\mcitedefaultendpunct}{\mcitedefaultseppunct}\relax
\EndOfBibitem
\bibitem[Blunt \latin{et~al.}(2019)Blunt, Thom, and Scott]{Blunt2019}
Blunt,~N.~S.; Thom,~A. J.~W.; Scott,~C. J.~C. Preconditioning and Perturbative
  Estimators in Full Configuration Interaction Quantum Monte Carlo. \emph{J.
  Chem. Theory Comput.} \textbf{2019}, \emph{15}, 3537\relax
\mciteBstWouldAddEndPuncttrue
\mciteSetBstMidEndSepPunct{\mcitedefaultmidpunct}
{\mcitedefaultendpunct}{\mcitedefaultseppunct}\relax
\EndOfBibitem
\bibitem[Eriksen \latin{et~al.}(2020)Eriksen, Anderson, Deustua, Ghanem, Hait,
  Hoffmann, Lee, Levine, Magoulas, Shen, Tubman, Whaley, Xu, Yao, Zhang, Alavi,
  Chan, Head-Gordon, Liu, Piecuch, Sharma, Ten-no, Umrigar, and
  Gauss]{Eriksen2020a}
Eriksen,~J.~J.; Anderson,~T.~A.; Deustua,~J.~E.; Ghanem,~K.; Hait,~D.;
  Hoffmann,~M.~R.; Lee,~S.; Levine,~D.~S.; Magoulas,~I.; Shen,~J.;
  Tubman,~N.~M.; Whaley,~K.~B.; Xu,~E.; Yao,~Y.; Zhang,~N.; Alavi,~A.; Chan,~G.
  K.-L.; Head-Gordon,~M.; Liu,~W.; Piecuch,~P.; Sharma,~S.; Ten-no,~S.~L.;
  Umrigar,~C.~J.; Gauss,~J. The Ground State Electronic Energy of Benzene.
  \emph{J. Phys. Chem. Lett.} \textbf{2020}, \emph{11}, 8922\relax
\mciteBstWouldAddEndPuncttrue
\mciteSetBstMidEndSepPunct{\mcitedefaultmidpunct}
{\mcitedefaultendpunct}{\mcitedefaultseppunct}\relax
\EndOfBibitem
\bibitem[Lee \latin{et~al.}(2020)Lee, Malone, and Reichman]{Lee2020}
Lee,~J.; Malone,~F.~D.; Reichman,~D.~R. The Performance of Phaseless
  Auxiliary-Field Quantum Monte Carlo on the Ground State Electronic Energy of
  Benzene. \emph{J. Chem. Phys.} \textbf{2020}, \emph{153}, 126101\relax
\mciteBstWouldAddEndPuncttrue
\mciteSetBstMidEndSepPunct{\mcitedefaultmidpunct}
{\mcitedefaultendpunct}{\mcitedefaultseppunct}\relax
\EndOfBibitem
\bibitem[Loos \latin{et~al.}(2020)Loos, Damour, and Scemama]{Loos2020}
Loos,~P.-F.; Damour,~Y.; Scemama,~A. The Performance of {CIPSI} on the Ground
  State Electronic Energy of Benzene. \emph{J. Chem. Phys.} \textbf{2020},
  \emph{153}, 176101\relax
\mciteBstWouldAddEndPuncttrue
\mciteSetBstMidEndSepPunct{\mcitedefaultmidpunct}
{\mcitedefaultendpunct}{\mcitedefaultseppunct}\relax
\EndOfBibitem
\bibitem[Rask and Zimmerman(2021)Rask, and Zimmerman]{Rask2021}
Rask,~A.~E.; Zimmerman,~P.~M. Toward Full Configuration Interaction for
  Transition-Metal Complexes. \emph{J. Phys. Chem. A} \textbf{2021},
  \emph{125}, 1598\relax
\mciteBstWouldAddEndPuncttrue
\mciteSetBstMidEndSepPunct{\mcitedefaultmidpunct}
{\mcitedefaultendpunct}{\mcitedefaultseppunct}\relax
\EndOfBibitem
\bibitem[Not()]{Note-1}
Computer architecture: Intel Xeon E5-2697v4 (Broadwell) hardware (36 cores @
  2.3 GHz, 3.56 GB/core).\relax
\mciteBstWouldAddEndPunctfalse
\mciteSetBstMidEndSepPunct{\mcitedefaultmidpunct}
{}{\mcitedefaultseppunct}\relax
\EndOfBibitem
\bibitem[Noga and Bartlett(1987)Noga, and Bartlett]{ccsdt_paper_1_jcp_1987}
Noga,~J.; Bartlett,~R.~J. {The Full CCSDT Model for Molecular Electronic
  Structure}. \emph{{J}. {C}hem. {P}hys.} \textbf{1987}, \emph{86}, 7041\relax
\mciteBstWouldAddEndPuncttrue
\mciteSetBstMidEndSepPunct{\mcitedefaultmidpunct}
{\mcitedefaultendpunct}{\mcitedefaultseppunct}\relax
\EndOfBibitem
\bibitem[Noga and Bartlett(1988)Noga, and
  Bartlett]{ccsdt_paper_1_jcp_1988_erratum}
Noga,~J.; Bartlett,~R.~J. {Erratum: The Full CCSDT Model for Molecular
  Electronic Structure}. \emph{{J}. {C}hem. {P}hys.} \textbf{1988}, \emph{89},
  3401\relax
\mciteBstWouldAddEndPuncttrue
\mciteSetBstMidEndSepPunct{\mcitedefaultmidpunct}
{\mcitedefaultendpunct}{\mcitedefaultseppunct}\relax
\EndOfBibitem
\bibitem[Scuseria and {Schaefer, III}(1988)Scuseria, and {Schaefer,
  III}]{ccsdt_paper_2_cpl_1988}
Scuseria,~G.~E.; {Schaefer, III},~H.~F. {A New Implementation of the Full CCSDT
  Model for Molecular Electronic Structure}. \emph{{C}hem. {P}hys. {L}ett.}
  \textbf{1988}, \emph{152}, 382\relax
\mciteBstWouldAddEndPuncttrue
\mciteSetBstMidEndSepPunct{\mcitedefaultmidpunct}
{\mcitedefaultendpunct}{\mcitedefaultseppunct}\relax
\EndOfBibitem
\bibitem[Oliphant and Adamowicz(1991)Oliphant, and
  Adamowicz]{ccsdtq_paper_1_jcp_1991}
Oliphant,~N.; Adamowicz,~L. {Coupled-Cluster Method Truncated at Quadruples}.
  \emph{{J}. {C}hem. {P}hys.} \textbf{1991}, \emph{95}, 6645\relax
\mciteBstWouldAddEndPuncttrue
\mciteSetBstMidEndSepPunct{\mcitedefaultmidpunct}
{\mcitedefaultendpunct}{\mcitedefaultseppunct}\relax
\EndOfBibitem
\bibitem[Kucharski and Bartlett(1992)Kucharski, and
  Bartlett]{ccsdtq_paper_2_jcp_1992}
Kucharski,~S.~A.; Bartlett,~R.~J. {The Coupled-Cluster Single, Double, Triple,
  and Quadruple Excitation Method}. \emph{{J}. {C}hem. {P}hys.} \textbf{1992},
  \emph{97}, 4282\relax
\mciteBstWouldAddEndPuncttrue
\mciteSetBstMidEndSepPunct{\mcitedefaultmidpunct}
{\mcitedefaultendpunct}{\mcitedefaultseppunct}\relax
\EndOfBibitem
\bibitem[Raghavachari \latin{et~al.}(1989)Raghavachari, Trucks, Pople, and
  Head-Gordon]{original_ccsdpt_paper}
Raghavachari,~K.; Trucks,~G.~W.; Pople,~J.~A.; Head-Gordon,~M. {A Fifth-Order
  Perturbation Comparison of Electron Correlation Theories}. \emph{{C}hem.
  {P}hys. {L}ett.} \textbf{1989}, \emph{157}, 479\relax
\mciteBstWouldAddEndPuncttrue
\mciteSetBstMidEndSepPunct{\mcitedefaultmidpunct}
{\mcitedefaultendpunct}{\mcitedefaultseppunct}\relax
\EndOfBibitem
\bibitem[Bomble \latin{et~al.}(2005)Bomble, Stanton, K{\'a}llay, and
  Gauss]{bomble_ccsdt_pq_jcp_2005}
Bomble,~Y.~J.; Stanton,~J.~F.; K{\'a}llay,~M.; Gauss,~J. {Coupled-Cluster
  Methods Including Noniterative Corrections for Quadruple Excitations}.
  \emph{{J}. {C}hem. {P}hys.} \textbf{2005}, \emph{123}, 054101\relax
\mciteBstWouldAddEndPuncttrue
\mciteSetBstMidEndSepPunct{\mcitedefaultmidpunct}
{\mcitedefaultendpunct}{\mcitedefaultseppunct}\relax
\EndOfBibitem
\bibitem[Not()]{Note-2}
Computer architecture used throughout the present study: Two Intel Gold 6130
  per node (32 cores @ 2.1 GHz, 5.53 GB/core). The initial base model
  calculation on the full system has not been included in all reported timings
  as this preliminary step is not amenable to embarrassingly parallel computing
  in the same way as MBE-FCI is.\relax
\mciteBstWouldAddEndPunctfalse
\mciteSetBstMidEndSepPunct{\mcitedefaultmidpunct}
{}{\mcitedefaultseppunct}\relax
\EndOfBibitem
\bibitem[Eriksen \latin{et~al.}(2014)Eriksen, Kristensen, Kj{\ae}rgaard,
  J{\o}rgensen, and Gauss]{eriksen_ccsd_pert_theory_jcp_2014}
Eriksen,~J.~J.; Kristensen,~K.; Kj{\ae}rgaard,~T.; J{\o}rgensen,~P.; Gauss,~J.
  {A Lagrangian Framework for Deriving Triples and Quadruples Corrections to
  the CCSD Energy}. \emph{{J}. {C}hem. {P}hys.} \textbf{2014}, \emph{140},
  064108\relax
\mciteBstWouldAddEndPuncttrue
\mciteSetBstMidEndSepPunct{\mcitedefaultmidpunct}
{\mcitedefaultendpunct}{\mcitedefaultseppunct}\relax
\EndOfBibitem
\bibitem[Eriksen \latin{et~al.}(2016)Eriksen, Kristensen, Matthews,
  J{\o}rgensen, and Olsen]{eriksen_convergence_ccpt_jcp_2016}
Eriksen,~J.~J.; Kristensen,~K.; Matthews,~D.~A.; J{\o}rgensen,~P.; Olsen,~J.
  {Convergence of Coupled Cluster Perturbation Theory}. \emph{{J}. {C}hem.
  {P}hys.} \textbf{2016}, \emph{145}, 224104\relax
\mciteBstWouldAddEndPuncttrue
\mciteSetBstMidEndSepPunct{\mcitedefaultmidpunct}
{\mcitedefaultendpunct}{\mcitedefaultseppunct}\relax
\EndOfBibitem
\bibitem[Eriksen \latin{et~al.}(2015)Eriksen, Matthews, J{\o}rgensen, and
  Gauss]{eriksen_quadruples_pert_theory_jcp_2015}
Eriksen,~J.~J.; Matthews,~D.~A.; J{\o}rgensen,~P.; Gauss,~J. {Communication:
  The Performance of Non-Iterative Coupled Cluster Quadruples Models}.
  \emph{{J}. {C}hem. {P}hys.} \textbf{2015}, \emph{143}, 041101\relax
\mciteBstWouldAddEndPuncttrue
\mciteSetBstMidEndSepPunct{\mcitedefaultmidpunct}
{\mcitedefaultendpunct}{\mcitedefaultseppunct}\relax
\EndOfBibitem
\bibitem[Eriksen \latin{et~al.}(2016)Eriksen, Matthews, J{\o}rgensen, and
  Gauss]{eriksen_open_shell_quadruples_jcp_2016}
Eriksen,~J.~J.; Matthews,~D.~A.; J{\o}rgensen,~P.; Gauss,~J. {Assessment of the
  Accuracy of Coupled Cluster Perturbation Theory for Open-Shell Systems. II.
  Quadruples Expansions}. \emph{{J}. {C}hem. {P}hys.} \textbf{2016},
  \emph{144}, 194103\relax
\mciteBstWouldAddEndPuncttrue
\mciteSetBstMidEndSepPunct{\mcitedefaultmidpunct}
{\mcitedefaultendpunct}{\mcitedefaultseppunct}\relax
\EndOfBibitem
\bibitem[Greiner \latin{et~al.}(2024)Greiner, Gianni, Nottoli, Lipparini,
  Eriksen, and Gauss]{Greiner2024}
Greiner,~J.; Gianni,~I.; Nottoli,~T.; Lipparini,~F.; Eriksen,~J.~J.; Gauss,~J.
  {MBE-CASSCF Approach for the Accurate Treatment of Large Active Spaces}.
  \emph{J. Chem. Theory Comput.} \textbf{2024}, \emph{20}, 4663\relax
\mciteBstWouldAddEndPuncttrue
\mciteSetBstMidEndSepPunct{\mcitedefaultmidpunct}
{\mcitedefaultendpunct}{\mcitedefaultseppunct}\relax
\EndOfBibitem
\end{mcitethebibliography}
\end{document}